\documentstyle[prb,aps]{revtex}
\begin{document}
\title{ Dynamic Polarization of the LiH Molecule in  Strong Light Field in
Anomalous-Dispersion Domain}
\author{ Shtoff A.V.;Dmitriev Yu.Yu.;Gusarov S.I. }

\address{St.Petersburg State University, Institute of Physics, Department of
Theoretical Physics, Oulianovskaya 1, Starii Petergof
198904 St.Petersburg Russia}
\date{\today}
\maketitle
\begin{abstract}

A new method is proposed to calculate the polarization vector of a molecule
in a monochromatic external field in the anomalous-despersion domain.
The method takes into account the instantaneous switching of the field.
A simple modification of the method allows one to consider a
more general switching procedure. As an illustration of the method Fourier
components of the polarization vector of the LiH molecule in the anomalous
-dispersion domain is calculated.

\end{abstract}

A new method of treating molecular polarization in monochromatic laser
field of high intensity is presented. This method is based on the Floquet
theory for molecular steady states in the extended Hilbert space,
introduced by H. Sambe \cite{ref_1}. It is combined with the general quantum
mechanical representation of molecular state in terms
of superposition of interacting configurations.
Neglecting the  molecular ionization and dicay we can restrict
the superposition to the finite number of terms.

 We choose the simplest form of time-periodic laser field potential
 $\widehat{V} (q,t)$ which is given by $\widehat{V}(q,t)=-\vec{F}
\vec{D}cos
(\omega t) $, where $\vec{D}$ is the electric dipole moment of the molecule,
$\vec{E}$ is the amplitude of the laser field which is considered to be
uniform. The general solution of the time-dependent Schr\"odinger equation
$$(\widehat{H}_{0} (q) + \widehat{V}(q,t)-i{\frac{ \partial }{\partial t} }
\Psi (q,t)=0 \eqno(1) $$ where $\widehat{H}_0 (q)$ is the time-independent
Hamiltonian
of unperturbed molecule is given by \cite{ref_2}:
$$
\Psi (q,t)= \sum\limits_j A_j exp(-i
{E}_j t) u_j(q,t) =
\sum\limits_{jns} A_j C_{ns}^{j} exp(-i( {E} -n \omega )t) \psi_s(q) \eqno(2)$$
 In this expression ${E}_j$ is quasi-energy of molecule, corresponding
to the steady-state eigenfunction $u_j (q,t) $, which is periodic in time $t$
with
the period
$ \frac{2 \pi }{\omega} $. The coefficients $C_{ns}^{j} $ can be obtained
within the variation principle and they are the eigenvectors of the
Hermitian matrix eigenvalue problem:
$$
\sum\limits_{sn} (\widehat{H}_{rm,sn}- {E}_j \delta_{rm,sn} )
C_{ns}^j =0.
$$
The elements of the matrix are given by the formula
$ H_{rm,sn}=(\varepsilon_r+ n \omega)\delta_{rs}\delta_{mn}+
{ \frac{1}{2} } \vec{F} \vec{D}_{rs} ( \delta_{m,n-1} + \delta_{m,n+1} )$.
In this formula $\varepsilon_{r}$ is  the energy of $r$-th stationary state
$ \psi_r$ of the unperturbed molecule, $\vec{D}_{rs}=
< \psi_r | \vec{D} | \psi_s> $. The coefficients $A_j$ are defined by initial
conditions. If the laser field is switched on
instantaneously ( which
can be considered as a good first approximation ) one finds that
$A_j$ satisty the set of the orthogonality conditions
$ \sum\limits_{j} B_{sj} A_{j} = \delta_{sk}$ where $B_{sj}= \sum\limits_{n}
C_{ns}^{j} $  and k is a quantum number of initial
molecular stationary state.
 Polarization vector of the molecule is calculated as the expectation value
$$
\vec{P}= < \Psi (q,t)| \vec{D} |\Psi (q,t) >.\eqno(3)
$$
 In case of the weak non-resonant field the wave function (2) consists of the
single
term of the steady state function $u_k (q,t)$ to which the wave function of the
molecule
evolves in the laser field. Therefore, in the non-resonant laser field
the polarization vector (3) depends on the time t periodically.

In strong resonant field
polarization vector, generally, is not periodic in time, it is because of the
terms
which are proportional to the phase factor $$ exp (-i (E_j - E_i) t), $$
This vector contains also a periodic in time component
$ \vec{P}_1 = \sum\limits_{j} A_j^2 <u_j |\vec{D} |u_j> $.
The periodic part of polarization vector can be expanded
in Fourier series : $ \vec{P}_1 = \sum\limits_n \vec{P}_{n \omega}
cos( \omega t )$, where Fourier components $ \vec{P}_{n \omega} $ are
$$
 \vec{P}_{n \omega}= - \sum\limits_{msr} \vec{D}_{sr}
 (2- \delta_{0n})C_{sm}^j C_{r,m+n}^{j}
$$
If the laser field is weak and non-resonant the polarization $ \vec{P}_{n
\omega} $
as a function of the field amplitude can be expanded in power series with
coefficients which determine susceptibilities (or polarizabilities) of the
molecule and describe its electric and optical properties.
As an example, we present results
of calculation of the first Fourier component $P_{1 \omega} $ of LiH molecule
dipole moment in the laser field which is parallel to the molecular axis.
The curves of  $P_{1 \omega} $ as a function of the field
amplitude $E$ is shown in Fig. 1.
They illustrate that function  $P_{1 \omega} $
depends strongly on frequencies of the laser field.
In the two-level approximation this function can be estimated by the
value $$P_{1 \omega} = -\frac{{D}_{12} (\omega - {\omega}_{12} )F }{
(\omega- {\omega}_{12})^2-{D}_{12}^2 E^2}, $$
where $\omega_{12}$ is the transition frequency.
It is obvious  that the power series which corresponds to this formula
converges only if $ E < |\frac {\omega-\omega_{12} }{ D_{12} }|$.
It means that for the strong resonant field the polarization
vector can not be expanded in power series of the field amplitude
and, therefore, the molecular properties can not be characterized by the set
of susceptibilities . In an alternative way one can consider the value
 $ \chi(\omega,E) =\frac{p_{1\omega}}{E} $
which depends on the field. This value is evidently equal to the polarizability
of the molecule for small values of $F$ and it is infinite if $\omega $ tends
to
$\omega_{12}$. In Fig.2 $\chi{\omega,E}$ is presented as a function of
frequency for two values of the field amplitude. One can easily see
that for both amplitudes which correspond to the strong field
 $ \chi(\omega,E) $ remaines finite in the anomalous-\~
dispersion domain. Hence, it can be considered as a field intensity
dependent polarizability of the molecule in the laser field.

The authors aknowledge Prof. M.N.Adamov for helful remarks and attention.


\end{document}